\documentclass[useAMS, usenatbib]{mn2e} 

\usepackage{graphicx,amssymb}

\title[Cosmic ray acceleration at Oblique Shocks]{Cosmic ray acceleration at oblique shocks}
\author[A.R. Bell, K.M. Schure and B. Reville]
{A. R. Bell\thanks{E-mail:t.bell1@physics.ox.ac.uk},
K.M. Schure and B. Reville\\
Clarendon Laboratory, University of Oxford, Parks Road, Oxford OX1 3PU, UK\\ }

\begin{document}

\newcommand\araa{{ARA\&A}}
\newcommand\apj{{ApJ}}
\newcommand{\apjl}{ApJL}
\newcommand\apjs{{ApJS}}
\newcommand\aap{{A\&A}}
\newcommand\mnras{{MNRAS}}
\newcommand\rmxaa{{Rev. Mexicana Astron. Astrofis.}}
\newcommand\nat{{Nature}}
\newcommand\physrep{{Phys.~Rep.}}
\newcommand\ssr{{Space Sci. Rev.}}
\newcommand\pasa{{ARA\&A}}
\newcommand\vv{{\bf v}}
\newcommand\fv{{\bf f}}
\newcommand\uv{{\bf u}}
\newcommand\Ev{{\bf E}}
\newcommand\Bv{{\bf B}}

\date{}
\pagerange{\pageref{firstpage}--\pageref{lastpage}}\pubyear{2011}
\maketitle
\label{firstpage}
\begin{abstract}
We show that the diffusion approximation 
breaks down for particle acceleration at oblique shocks with velocities typical of young supernova remnants. 
Higher order anisotropies flatten the spectral index at quasi-parallel shocks and steepen the spectral index 
at quasi-perpendicular shocks.  We compare the theory with observed spectral indices.
\end{abstract}

\section{Introduction}

The well-established theory of cosmic ray (CR) acceleration by strong non-relativistic
shocks predicts a differential energy power law spectrum $n(E)dE\propto E^{-s}dE$ 
with a spectral index $s=2$ (Axford, Leer \& Skadron 1977, Krimskii 1977, Bell 1978, Blandford \& Ostriker 1978).  
However, a variety of spectral indices
is observed.  Galactic CR have a steeper spectrum even when
allowance is made for energy-dependent propagation losses (Hillas 2005).  Steeper spectra
are also observed in young supernova remnants (SNR) as discussed below.  
Deviations from a $s=2$ spectrum are well known for relativistic shocks
(for example: 
        Kirk \& Heavens 1989,
        Ostrowski 1991,
        Baring Ellison \& Jones 1993,
        Ruffalo 1999,
        Gieseler et al 1999,
        Kobayakawa, Honda \& Samura 2002,
        Achterberg et al 2001).
Here we show that corrections to theory produce a spectral index 
that deviates significantly from $s=2$
even at shock velocities as low as 10,000 km s$^{-1}$ typical of young supernova remnants.
The magnitude of the effect depends on the ratio of the velocity of the accelerating particle to the shock velocity.  
Hence, it may also apply to the acceleration of non-relativistic particles at
heliospheric shocks.
The spectrum can either steepen or flatten depending on the angle between the shock normal and the
large scale upstream magnetic field. Quasi-perpendicular shocks (Jokipii 1982, 1987) show particularly strong steepening. 
The effect is small for parallel shocks where changes in the spectral index are second order in 
the ratio of the particle velocity to the shock velocity.
Non-linear hydrodynamic feedback of the CR pressure onto the shock structure can also change the
spectral index (Drury 1983, Dury \& V\"{o}lk 1981, Ellison, Baring \& Jones 1996) 
but we do not take account of this in the present paper.  
We assume that the fluid velocity is uniform both upstream and downstream of the shock.  
A further assumption lying at the heart of our method is that the magnetic field
can be divided into (i) a large scale field that is uniform apart from a change of direction and magnitude at the shock and (ii) a
small scale field that fluctuates on scales smaller than the CR Larmor radius and deflects the
CR trajectories through small angles to cumulatively scatter the CR.
This ignores magnetic field line wandering which can also change the CR spectral index 
(Kirk, Duffy \& Gallant 1996).
In our calculations we further assume that the fluid is always compressed by a factor of four at the shock, 
which neglects the increase in compression at the shock due to the different equation of state 
 of the relativistic component (ratio of specific heats = 4/3) and the effects of finite Mach number.  
With these assumptions we isolate and concentrate on the breakdown of the diffusive 
approximation and the effect of high order anisotropies in the CR distribution near the shock.

\section{Equations}
Cosmic rays are thought to be accelerated to
high energy by crossing back and forth across a shock in a first order Fermi process, 
with a fractional energy
gain of order $u_s/c$ at each crossing where $u_s$ is the shock velocity.  
A small proportion of the CR injected at low energy proceed to cross the shock
many times before escaping the shock environment. Statistically this results
in a power law stretching up to a few PeV in the galaxy and probably approaching
ZeV extragalactically.  Here we consider CR protons of mass $m_p$, but the analysis applies
equally well to heavier ions or to electrons.  The shape and extent of the CR spectrum is described
by the CR distribution function $f(x,{\bf p},t)$ in momentum ${\bf p}$.  
We locate a planar shock at $x=0$.  Upstream of the shock ($x<0$) the ambient plasma
flows towards the shock at the shock velocity $u_s$.  For a strong non-relativistic shock
the plasma  velocity is reduced to $u_s/4$ in the downstream plasma.  The
equation for the CR distribution function is
$$
\frac{\partial f}{\partial t}+(v_x+u)\frac{\partial f}{\partial x}  
-\frac{\partial u}{\partial x} p_x\frac{\partial f}{\partial p_x}
-\frac{u}{c}\frac{\partial u}{\partial x} p\frac{\partial f}{\partial p_x}
+e{\bf v}\wedge{\bf B}.\frac{\partial f}{\partial {\bf p}}
=C(f)
\eqno{(1)}
$$
where $f$ is defined in the local fluid rest frame moving at velocity $u(x)$ in the $x$ direction
and relativistic effects due to $\gamma_u=(1-u^2/c^2)^{-1/2}\neq 1$ have been neglected.
$v_x$ is the $x$-component of the CR velocity ${\bf v}$, 
 $\bf B$ is the local large scale magnetic field, and $C(f)$ represents CR scattering by small scale fluctuations in the magnetic field. In the local fluid rest frame, $C(f)$ does not change the CR energy, but merely deflects its trajectory by diffusion in angle.  $\partial u/\partial
x$ is zero everywhere except at the shock where it produces CR acceleration.
Building on our experience of modelling laser-plasmas
we solve the Vlasov-Fokker-Planck (VFP) equation by expressing the CR distribution function as a sum
of spherical harmonics (Bell et al 2006).
\begin{eqnarray*}
f(x,{\bf p},t)=\sum _{l,m} f_l^m(x,p,t)P_l^{|m|}(\cos \theta)e^{im\phi}\\
l=0,\infty \ \ \ \ m=-l,l\ \  \ \ f_l^{-m}=\left( f_l^m \right)^*
\end{eqnarray*}
where $p_x=p\cos \theta$, $p_y=p\sin \theta \cos \phi$ and $p_z=p\sin \theta \sin \phi$.  
The VFP equation can be separated into individual moment equations for the evolution of $f_l^m(x,p,t)$:
\begin{eqnarray*}
\frac{\partial f_l^m}{\partial t}+u \frac{\partial f_l^m}{\partial x}
+c\left[ \frac{l-m}{2l-1}\frac{\partial f_{l-1}^m}{\partial x}
+\frac{l+m+1}{2l+3}\frac{\partial f_{l+1}^m}{\partial x}\right] 
 \\ 
-im\frac{ceB_x}{p}f_l^m
+\frac{ceB_z}{2p}\beta_l^m
\\
 -p\frac{\partial u}{\partial x}\bigg[
\frac{(l-m)(l-m-1)}{(2l-3)(2l-1)}
     \left( \frac{\partial f_{l-2}^m}{\partial p}-(l-2)\frac{f_{l-2}^m}{p}\right)
 \\
+\frac{(l-m)(l+m)}{(2l-1)(2l+1)}
     \left( \frac{\partial f_l^m}{\partial p}+(l+1)\frac{f_l^m}{p}\right)
 \\
+\frac{(l-m+1)(l+m+1)}{(2l+1)(2l+3)}
     \left( \frac{\partial f_l^m}{\partial p}-l\frac{f_l^m}{p}\right)
\\
+\frac{(l+m+1)(l+m+2)}{(2l+3)(2l+5)}
     \left( \frac{\partial f_{l+2}^m}{\partial p}+(l+3)\frac{f_{l+2}^m}{p}\right)
     \bigg]
\\
 -p\frac{u}{c}\frac{\partial u}{\partial x}\bigg[
\frac{l-m}{2l-1}
     \left( \frac{\partial f_{l-1}^m}{\partial p}-(l-1)\frac{f_{l-1}^m}{p}\right)
\\
+\frac{l+m+1}{2l+3}
     \left( \frac{\partial f_{l+1}^m}{\partial p}+(l+2)\frac{f_{l+1}^m}{p}\right)
     \bigg]
\\
=-\frac{l(l+1)}{2}\nu f_l^m
\end{eqnarray*}
$$\eqno{(2)}
$$
where $\beta_l^m=(l-m)(l+m+1)f_l^{m+1}-f_l^{m-1}$ for $m>0$ and $\beta_l^0=2l(l+1)\Re{(f_l^1)}$.
The collision term $C(f)$ on the right hand side reduces to an algebraic form because 
spherical harmonics are eigenfunctions of the Laplacian operator representing
diffusion  in $\theta$ and $\phi$.
The co-ordinate system has been chosen such that $B_y=0$ and $\nu(x)$
is the collision frequency. 
The first three moment equations $(l=0,1)$ are
 \begin{eqnarray*}
\frac{\partial f_0^0}{\partial t}
+u\frac{\partial f_0^0}{\partial x}
+\frac{c}{3}\frac{\partial f_1^0}{\partial x}
-\frac{1}{3}\frac{\partial u}{\partial x}  p\frac{\partial f_0^0}{\partial p}
\hskip 3 cm \\
-\frac{2}{15}\frac{\partial u}{\partial x}\left(p\frac{\partial f_2^0}{\partial p}+3f_2^0\right)
-\frac{1}{3} \frac{u}{c} \frac{\partial u}{\partial x}\left(p\frac{\partial f_1^0}{\partial p}+2f_1^0\right)
=0
\end{eqnarray*}
\begin{eqnarray*}
\frac{\partial f_1^0}{\partial t}
+c\frac{\partial f_0^0}{\partial x}
+2 \omega_z \Re{(f_1^1)}
+\nu f_1^0
+u\frac{\partial f_1^0}{\partial x}
+\frac{2c}{5}\frac{\partial f_2^0}{\partial x}
\hskip 0.8 cm \\
-\frac{\partial u}{\partial x}\left(\frac{3p}{5}\frac{\partial f_1^0}{\partial p}+\frac{1}{15}f_1^0 + \frac{6p}{35}\frac{\partial f_3^0}{\partial p}+\frac{24}{35}f_3^0\right)
\\
-\frac{u}{c}\frac{\partial u}{\partial x}
\left ( p\frac{\partial f_0^0}{\partial p}
+\frac{2p}{5} \frac{\partial f_2^0}{\partial p}+\frac{6}{5}f_2^0\right)=0
\end{eqnarray*}
\begin{eqnarray*}
\frac{\partial f_1^1}{\partial t}
-i \omega _x f_1^1
-\frac{\omega_z}{2}\Re{(f_1^0)}
+\nu f_1^1
\hskip 4 cm \\
+u\frac{\partial f_1^1}{\partial x}
+\frac{3c}{5}\frac{\partial f_2^1}{\partial x}
-\frac{\partial u}{\partial x}\left(\frac{p}{5}\frac{\partial f_1^1}{\partial p}-\frac{1}{5}f_1^1\right)
\hskip 2.1 cm\\
-\frac{\partial u}{\partial x}\left(\frac{12p}{35}\frac{\partial f_3^1}{\partial p}+\frac{48}{35}f_3^1\right)
-\frac{u}{c}\frac{\partial u}{\partial x}
\left ( \frac {3p}{5}\frac{\partial f_2^1}{\partial p}+\frac{9}{5}f_2^1\right)=0
\end{eqnarray*}
$$\eqno{(3)}
$$
where $\omega _z =ceB_z/p$ and $\omega _x =ceB_x/p$.
Supernova ejecta initially expand at velocities up to $\sim c/5$, and the SNR shock
velocity decreases to $\sim c/100$ during the first 1000 years.   
In the limit of small $u/c$ the $l=0$ and $l=1$ equations dominate.
In the Chapman-Enskog expansion, valid for scalelengths much larger than the CR scattering length, 
the magnitude of $f_l^m$ decreases with increasing $l$, such that 
$f_l^m \sim (u/c)^lf_0^0$.  
With this expansion, $f_l^m$ can be neglected for $l > 1$, in which case the above three equations ($l=0$ and $l=1$) reduce to
$$
\frac{c}{3}\frac{\partial f_1^0}{\partial x}
+u\frac{\partial f_0^0}{\partial x}
=0
$$
$$
c\frac{\partial f_0^0}{\partial x}
+2\omega _z \Re{(f_1^1)}
=-\nu f_1^0
$$
$$
-i \omega _x f_1^1
-\frac{\omega_z}{2}f_1^0
=-\nu f_1^1
\eqno{(4)}
$$
where $f_0^0$ and $f_1^0$ are real.  The term $u\partial f_0^0/\partial x$ 
in the first equation cannot be neglected as small to order
O$(u/c)$ since $f_0^0 \sim (c/u) f_1^0$, but terms involving $f_2^m$ are neglected since they are of order $(u/c)f_1^m$.  
$f_1^m$ can be eliminated between the equations 
resulting in a diffusion equation for $f_0^0$.
$\partial u/\partial x$ is non-zero only at the shock, so the upstream diffusive solution 
of these equations for the CR density $f_0^0$ for a general oblique shock is
$$
f_0^0(x)=f_0^0(0) \exp (\kappa x)
\hskip 0.5 cm {\rm where} \hskip 0.5 cm
\kappa = \frac{3\nu u_s}{c^2}  \frac{\nu^2+\omega_x^2+\omega_z^2}{\nu^2+\omega_x^2}
\eqno{(5)}
$$ 
The corresponding CR fluxes are 
$$
f_1^0(x)=-\frac{3u_s}{c}f_0^0(x)
\hskip 0.25 cm {\rm and} \hskip 0.25 cm
 f_1^1(x)=-\frac{3u_s}{2c}\frac{(\nu+i\omega_x)\omega_z}{\nu^2+\omega_x^2}f_0^0(x)
 \eqno{(6)}
 $$
An alternative tensor notation (Johnston 1960)  
 expresses the CR distribution function to first order in $u/c$ as
 $f=f_0+{\bf f_1}.{\bf p}/p $.  
 The tensor expansion is related to the spherical harmonic expansion by $f_x=f_1^0$, $f_y=-\Re\{ 2f_1^1\}$, and  $f_z=\Im\{ 2f_1^1\}$.
 We choose the $y$ and $z$ directions such that $B_y=0$.
 In this geometry, $f_x$ is parallel to the shock normal, $f_z$ is perpendicular to the shock normal and lies in the plane 
 containing the magnetic field and the shock normal, and $f_y$  is perpendicular to both the shock normal and the magnetic field.
  $f_1^0$, or $f_x$, represent CR diffusion parallel to the shock normal. 
 CR drift along the magnetic field is represented by a combination of $f_1^0$ and $\Im\{f_1^1\}$.  
 Both real and imaginary parts of $f_1^1$ are zero everywhere for a parallel shock. 
 
Downstream of the shock ($x>0$), the corresponding solution to the same equations is that for a uniform isotropic CR distribution at rest
with respect to the background plasma: $f_0^0(x)=f_0^0(0)$, $f_1^0=f_1^1=0$ where $f_0^0(0)$ is the CR density at the shock.  
Matching the upstream
and downstream CR distribution functions at the shock requires $f_0^0$ to
be continuous at the shock.  Integrating the $(l=1,m=0)$ moment equation
across the shock requires $f_1^0-(u/c)p\partial f_0^0/\partial p$ to be continuous
at the shock.  The term $-(u/c)p\partial f_0^0/\partial p$ arises from the frame 
transformation across the shock. 
Assuming a strong shock
in which $u=u_s/4$ downstream, the upstream flux 
$f_1^0-(u_s/c)p\partial f_0^0/\partial p$ must equal the downstream flux $-(u_s/4c)p\partial f_0^0/\partial p$, 
giving $p\partial f_0^0/\partial
p=-4f_0^0$ from equation (6).  This leads to the well-known power law $f_0^0\propto p^{-4}$, equivalent to 
$n(E)\propto E^{-2}$ for CR acceleration
at a strong shock.  
This solution is correct in the limit of low shock velocity 
for a parallel shock in which $B_y=B_z=0$.  

The $f_0^0$ and $f_1^m$ components of the distribution function can be matched across a parallel shock.  However, it is not possible to match the solution
for $f_1^1$ across an oblique shock when the higher order terms ($f_l^m$ for $l > 1$) are neglected.
The Chapman-Enskog expansion breaks down because of the abrupt gradient at the shock.  
In the limit of small shock velocity, the mismatch at an oblique shock
is unimportant and introduces only a correction of order $u/c$ to the solution. 
Consequently, the spectrum at an oblique shock is still  $f_0^0\propto p^{-4}$ in the low velocity limit, 
but the corrections of order
$u/c$ become important at the expansion velocities of young SNR.  
The corrections are more important for oblique shocks than for parallel shocks for which the 
corrections are of  order $(u/c)^2$.  
As we show below, the CR spectrum is expected to deviate from $p^{-4}$ more strongly for oblique shocks than for parallel shocks.

One reason why oblique shocks deviate more strongly from a $p^{-4}$ spectrum is that the expansion parameter 
is more correctly $u/ (c\cos \theta) $ rather than $u/c$, where $\theta $ is the angle between the magnetic field and 
the shock normal.
CR diffuse away from the shock along magnetic field lines at the angle
$\theta $ to the shock normal.  If $u > c\cos \theta $, it is impossible for CR to escape upstream of the shock 
by streaming along field lines.  Shocks with a velocity
$u$ exceeding $c\cos \theta $  are accounted `superluminal', even if the shock velocity is in fact considerably less than the speed of light.
Corrections to the spectral index $\gamma$, where $f_0^0(0)\propto p^{-\gamma}$, are especially strong when $u/\cos \theta$ approaches $c$.  

The aim of this paper is to calculate the CR spectral index at oblique shocks at shock velocities relevant to young SNR.  
Integration of the first of equations (3) from far upstream to far downstream gives
$$
\gamma =\frac{
 \int _{-\infty}^{\infty}\left (3f_0^0+6f_2^0/5+2uf_1^0/c\right ) (\partial u/\partial x) dx
 -3u(\infty)f_0^0(\infty)}
{\int _{-\infty}^{\infty}\left ( f_0^0 +2f_2^0/5 +uf_1^0/c \right ) (\partial u/\partial x) dx}
 \eqno{(7)}
$$
where $u(\infty)$ and $f_0^0(\infty)$ are the values of $u$ and $f_0^0$ far downstream.  
At the shock $f_0^0(0)\propto p^{-\gamma}$.
To leading order in $u/c$, and for a discontinuous shock transition
$$
\gamma =3+\frac{3u(\infty) }{u_s - u(\infty)}\frac{f_0^0(\infty) }{f_0^0(0)}
 \eqno{(8)}
$$
where $f_0^0(0)$ is the value of $f_0^0$ at the shock.  For a shock compression of 4 ($u(\infty)=u_s/4$), 
$\gamma = 4$ since the downstream CR density is uniform ($f_0^0(\infty)=f_0^0(0)$) to order $u/c$.
We shall show that for an oblique shock, $f_0^0(\infty)\neq f_0^0(0)$, and $\gamma$
differs from 4 to order $u/c$.  

The ratio ${f_0^0(\infty) }/{f_0^0(0)}$ is important for determining the spectral index $\gamma$ because $f_0^0(0)$ 
determines the rate at which CR are accelerated by crossing and recrossing the shock, while $f_0^0(\infty)$
determines the rate at which CR are advected away from the shock downstream. The spectral index is determined 
by the number of times CR cross the shock before escaping downstream.  
Hence $\gamma$ depends on  ${f_0^0(\infty) }/{f_0^0(0)}$, 
although the values of $f_1^0$ and $f_2^0$ at the shock can also change the spectral index to higher order
as evident in equation (7).

\section{Numerical Model}

For all except the highest energy CR, the acceleration time is shorter than the time for which the shock exists, 
so we seek a time-independent solution with $\partial f_l^m/\partial t=0$  for all $l$ and $m$.
We assume that the collision frequency $\nu$ due to CR scattering by small scale magnetic field 
is proportional to the local CR Larmor frequency:
$$ \nu (x) \propto |{\bf B}(x)| \eqno{(9)}$$
where ${\bf B}$ is the local magnetic field, thus making the ratio of the 
CR mean free path to the Larmor radius the same both upstream and downstream of the shock.
By definition of $\nu$ as a frequency for small angle scattering, the coefficients $f_l^m$ decay exponentionally 
at a rate $[l(l+1)/2]\nu$ in the absence of all other effects.
The large-scale magnetic field is assumed to be uniform in the upstream plasma, 
$B_x =B_0 \cos \theta$, 
$B_y=0$, $B_z =B_0 \sin \theta$ and compressed by the shock to give a uniform downstream field,
$B_x =B_0 \cos \theta$, 
$B_y=0$, $B_z =4 B_0 \sin \theta$.
As already stated, we assume throughout that the plasma is compressed by a factor 4 at the shock, 
even at high shock velocity where the compression is in reality increased
by the reduced relativistic ratio of specific heats. 

From the self-similar structure of equation (2) it can be shown that the steady state CR distribution function is 
a power law in momentum at the shock.
The shock is the only point in space at which $\partial u /\partial x$ is non-zero, and hence the only point at which $f(x,p)$ 
is connected to the values of $f(x,p+\Delta p)$ and   $f(x,p-\Delta p)$.  
If the spectral index at the shock is $\gamma$, 
$p\partial f_l^m/\partial p$ can be replaced by $-\gamma f_l^m$ in equation (2) so that 
it becomes an equation with differentials in $x$ but not $p$.
$p$ only appears in combination with the magnetic field as $eB_x/p$ and $eB_z/p$ as the inverse of the CR Larmor radius
by which all distances are scaled.
The objective of the calculation is then to find the value of $\gamma$ for which the 
boundary conditions are $f_l^m=0$ for all $l,m$  
far upstream 
where $x\rightarrow - \infty$, and for which only $f_0^0$ is non-zero far downstream. 
Downstream, $f_l^m \rightarrow 0$ as $x\rightarrow  \infty$ for $l >0$.

The steady state VFP equation is non-linear because of the multiplication of the two unknowns $f$ and $\gamma$ 
in terms proportional to $p\partial f_l^m/\partial p=-\gamma f_l^m$.  
Making the replacement $p\partial f_l^m/\partial p=-\gamma f_l^m$, we
iterate to a steady state solution by restoring the time-derivative in equation (2), 
fixing the value of $f_0^0$ at the shock,
and improving the value of $\gamma$ as equation (2)
is integrated forward in time until a steady state is reached 
in which CR advect at constant density through a computational boundary placed far downstream of the shock.
During the time integration, $\gamma$  is calculated at each timestep
from equation (7).
Equation (2) is solved in finite difference form with $f_l^m$ defined at the cell centres for even
$l$, and defined on the cell boundaries for odd $l$.  
All spatial derivatives are then naturally centre-differenced.
A stretched spatial grid is used to give fine resolution near the shock, where $f_l^m$ changes rapidly on scales less than a 
CR Larmor radius.  A coarser resolution is used at large distances from the shock where $f_l^m$ changes on the 
scalelength of the shock precursor.  
The background fluid velocity $u$ is pre-determined and changes smoothly by a factor of four across a thin shock transition.
$\bf B$ is determined by the flux freezing condition.  
At the shock, we impose the form
$u=0.625u_s -0.375u_s \tanh (x/x_s)$.  For high shock velocity $u_s=c/5$, the lengthscale $x_s$ for the shock transition 
is set to 0.01 times the upstream CR Larmor radius, and the spacing of the computational grid at the shock is set 
to $\Delta x = x_s/3$.  
As shown in figure 1(a), halving $\Delta x$ makes very little difference in the case requiring 
the highest resolution (a perpendicular shock with $u_s=c/5$).
$x_s$ and $\Delta x$ are set a factor of two larger for low shock velocities where the natural scalelengths are larger.  For quasi-perpendicular high velocity shocks ($u_s=c/5$), the spherical harmonic expansion was extended to $l=11$.  Test calculations with a larger number, $l=15$, made negligible difference to the results.

With our neglect of non-linear hydrodynamic effects due to the CR pressure in the precursor, the shock transition takes place over 
a distance very much less than the Larmor radius of high energy CR.  
For this reason our chosen scalelength $x_s$ is much smaller than the Larmor radius.  However we cannot treat 
the shock as a discontinuity since $\partial u/\partial x$ would then be infinite which is not possible in our numerical treatment.
An alternative numerical method would be to treat the shock as a discontinuity with a matrix relating the values of $f_l^m$ each side of the shock,
$f_{l_1}^m=A_{l_1 l_2}^m f_{l_2}^m$. 
$A_{l_1 l_2}^m$ would be a full matrix in $(l_1,l_2)$ with the spherical harmonic expansion extended
to very high order.
A discontinuous shock introduces high order harmonics, making this approach unrealistic.
Smoothing the shock over a small distance $x_s$ terminates the harmonic expansion without significantly changing the spectral index.
However, as an additional check on the overall accuracy of the code (Code A), 
we wrote a 
completely separate code (Code B) which treats the shock jump differently, allowing a sharper transition at the cost of dispensing with second order terms in $u/c$ at the shock. 
Code B solves the reduced equations
\begin{eqnarray*}
\frac{\partial f_l^m}{\partial t}+u \frac{\partial f_l^m}{\partial x}
+c\left[ \frac{l-m}{2l-1}\frac{\partial f_{l-1}^m}{\partial x}
+\frac{l+m+1}{2l+3}\frac{\partial f_{l+1}^m}{\partial x}\right] 
 \\ 
-im\frac{ceB_x}{p}f_l^m
+\frac{ceB_z}{2p}\beta_l^m
-S_l^m
=-\frac{l(l+1)}{2}\nu f_l^m
\end{eqnarray*}
where
$$
S_0^0=\frac{1}{3}\frac{\partial u}{\partial x} p \frac{\partial f_0^0}{\partial p}
\hskip 0.3cm {\rm and} \hskip 0.3cm 
S_2^0=\frac{2}{3}\frac{\partial u}{\partial x} p \frac{\partial f_0^0}{\partial p}
 \eqno{(10)}
$$
and all other $S_l^m=0$.
$S_2^0$ must be included because it balances the jump in $f_1^0$ across the shock.
All other terms involving $\partial u/\partial x$ are smaller to order $u/c$ but not negligible.
Their omission substantially simplifies the numerical integration across the shock. Some minimal
smoothing of the shock transition on the scale of the computational grid is 
still included to allow the spherical harmonic expansion to be terminated.
The output from the code B is compared with the results from the main code (code A) in figure 2(b) for $\nu/\omega_g=0.1$ and $u/c=0.1$, which we consider to be a representative `standard' set of parameters.
As expected, the comparison shows a difference in the spectral
index due to the neglect of second order terms in $u/c$, but the two curves
are sufficiently close to give confidence that the results from Code A can be relied upon.

\section{Results}

\begin{figure*}
\begin{center}
\includegraphics[angle=0,width=16cm]{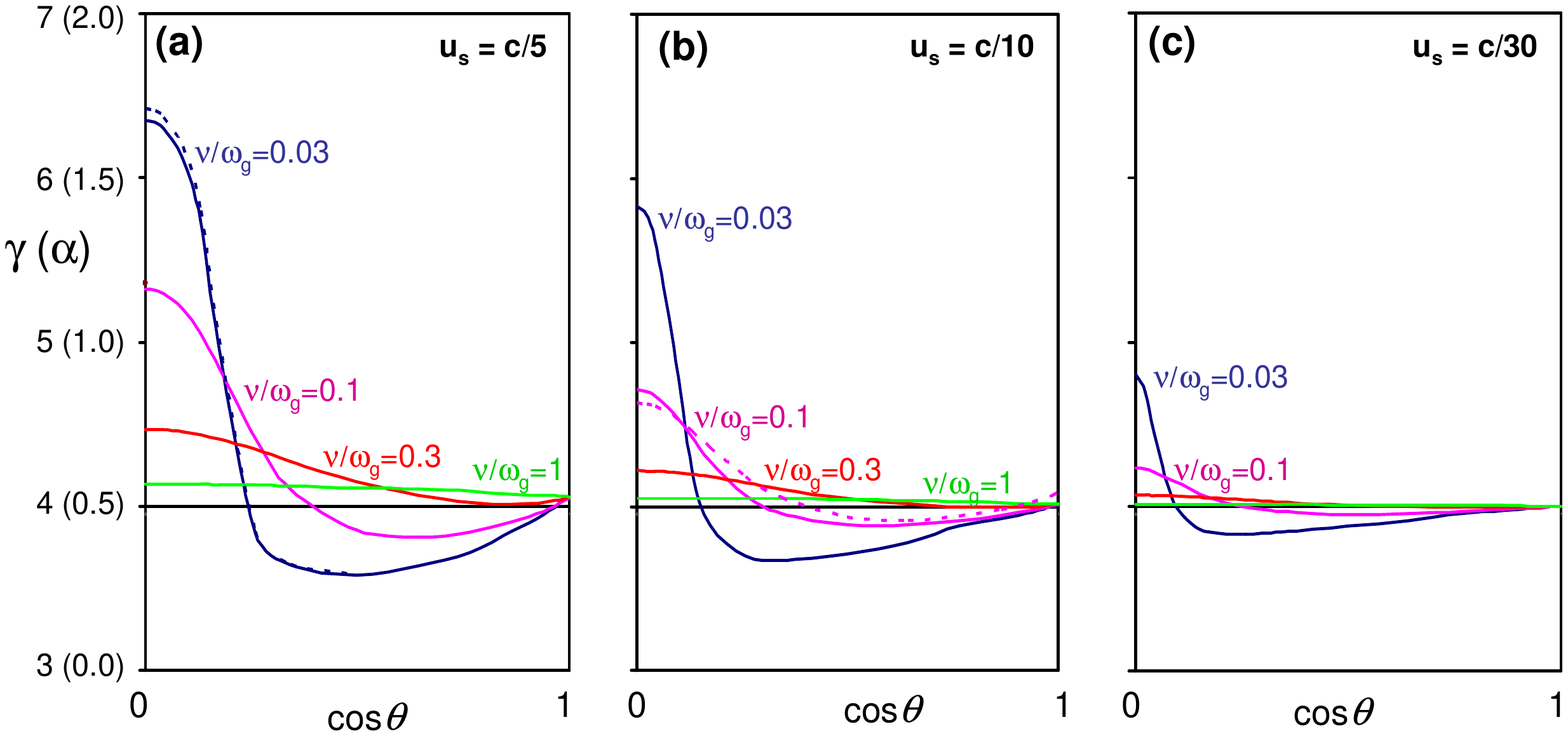}
\vskip 2 cm 
\caption{Spectral indices at three shock velocities (c/5, c/10, c/30) and four CR scattering rates
($\nu/\omega_g =$ 0.03, 0.1, 0.3 and 1.0).  $\gamma$ is the CR energy spectral index. $\alpha$ is the corresponding electron synchrotron spectral index, $\alpha=(\gamma-3)/2$.  $\theta$ is the shock obliquity.  At a perpendicular shock $\cos \theta =0$.  At a parallel shock $\cos\theta=1$. The dotted line in 1(a) next to the curve for $\nu/\omega_g=0.03$ is a calculation with finer spatial resolution with $\Delta x$ halved as an accuracy check on the most demanding calculation.  The dotted line in 1(b) next to the curve for $\nu/\omega_g=0.1$ is a comparison calculation with code B.  All other curves are calculated with Code A.}
\label{fig:Fig1}
\end{center}
\end{figure*}

Equation (2) was solved for a variety of shock velocities, $u_s=c/5,\ c/10\  \&\  c/30$,
and a variety of collisionalities, $\nu/\omega_g=0.03, \ 0.1, \ 0.3\ \& \ 1.0$, where $\omega _g=ceB_0/p$ is the CR Larmor frequency in 
the upstream plasma.
The spectral indices $\gamma$ are plotted in figure 1 as a function of $\cos \theta$
where $\theta$ is the angle between the shock normal and the magnetic field.
At a perpendicular shock $\cos \theta =0$.  At a parallel shock $\cos\theta=1$.
In the case of a shock advancing into a randomly disordered magnetic field $\cos \theta$ is uniformly distributed between 0 and 1.
Also plotted on the vertical axis is the synchrotron spectral index $\alpha=(\gamma
-3)/2$.

The results can be summarised as follows:
\newline
\noindent
1) In the case of parallel shocks, the CR spectral index deviates little from $\gamma=4$ ($\alpha=0.5$).
\newline
2) The quasi-parallel spectrum is flattened ($\gamma<4$, $\alpha <0.5$).
\newline
3) The quasi-perpendicular spectrum is steepened ($\gamma>4$, $\alpha >0.5$) and CR are accelerated less efficiently.
\newline
4) Deviation from $\gamma = 4$ is strongest for high shock velocity, 
but significant deviations are still present at velocities as low as 10,000 km sec$^{-1}$.
\newline
5) Deviations from $\gamma = 4$ are greater if the angular scattering time is much longer
than the CR Larmor gyration time ($\nu \ll \omega _g$).
\newline
6) In the case of a shock advancing into a magnetic field which is disordered on 
scales both larger and smaller than a CR Larmor radius, the spectral index averages to 
 $\gamma = 4$ since the mean value of $\gamma$ averaged over $\cos \theta$
is close to $\gamma = 4$ (see figure 1).  
\newline
7) The overall spectral index can only be strongly flattened if the shock is
predominantly quasi-parallel.  
Conversely, it can only be strongly steepened if the shock
is predominantly quasi-perpendicular.
\vskip 0.2 cm

The magnitude of spectral index deviation from $\gamma=4$ depends 
strongly on the ratio of the CR scattering frequency $\nu$ to the CR
Larmor freqency $\omega _g$.  The parameter $\nu/\omega _g$ is poorly known
both observationally and theoretically, but there are reasons for assuming 
that for SNR shocks $\nu/\omega _g$ lies in the band of 
values we choose for figure 1.
Very low collision frequencies ($\nu/\omega _g \ll 1$) can be ruled out for SNR shocks because CR acceleration
would then be too slow for CR acceleration to PeV energies as required by the galactic CR spectrum.
At the opposite limit, large collision frequencies ($\nu/\omega _g>1$) can probably be ruled out
since $\nu/\omega _g>1$ would correspond to fields with no large scale structure.  This is unlikely because some form of power-law magnetic field turbulence spectrum can be expected
which includes fluctations at wavenumbers less than, as well as greater than, the inverse of any particular CR Larmor radius. 
As a guess, and it is little more than that at present, it seems reasonable to suppose that
anisotropic CR drifts ($f_x$, $f_y$ \& $f_z$) might decay by a factor of two in the time it takes for
a CR to make one gyration through an angle of $2\pi$.  
This corresponds to $\nu/\omega _g=0.11$.
It is often presumed that `Bohm diffusion' applies to CR scattering, but this can variously 
and ambiguously be taken to mean $\nu/\omega _g=1$, $\nu/\omega _g=1/3$ 
or $\nu/\omega _g=0.125$ as in the NRL Plasma Formulary (Huba 1994).
$\nu/\omega _g=1$ would seem to be too large because
drift anisotropies then decay by a factor of $\sim \exp(-2\pi)=10^{-3}$ during one
CR gyration time, which would only be possible if the magnetic field responsible for 
small scale scattering is much greater than the field structured 
on the scale of a CR Larmor radius.
The value of the ratio $\nu/\omega _g$ is a major uncertainty
in our discussion, and we take $\nu/\omega _g=0.1$ as a `standard' value.

\begin{figure*}
\begin{center}
\vskip 2 cm 
\includegraphics[angle=0,width=12cm]{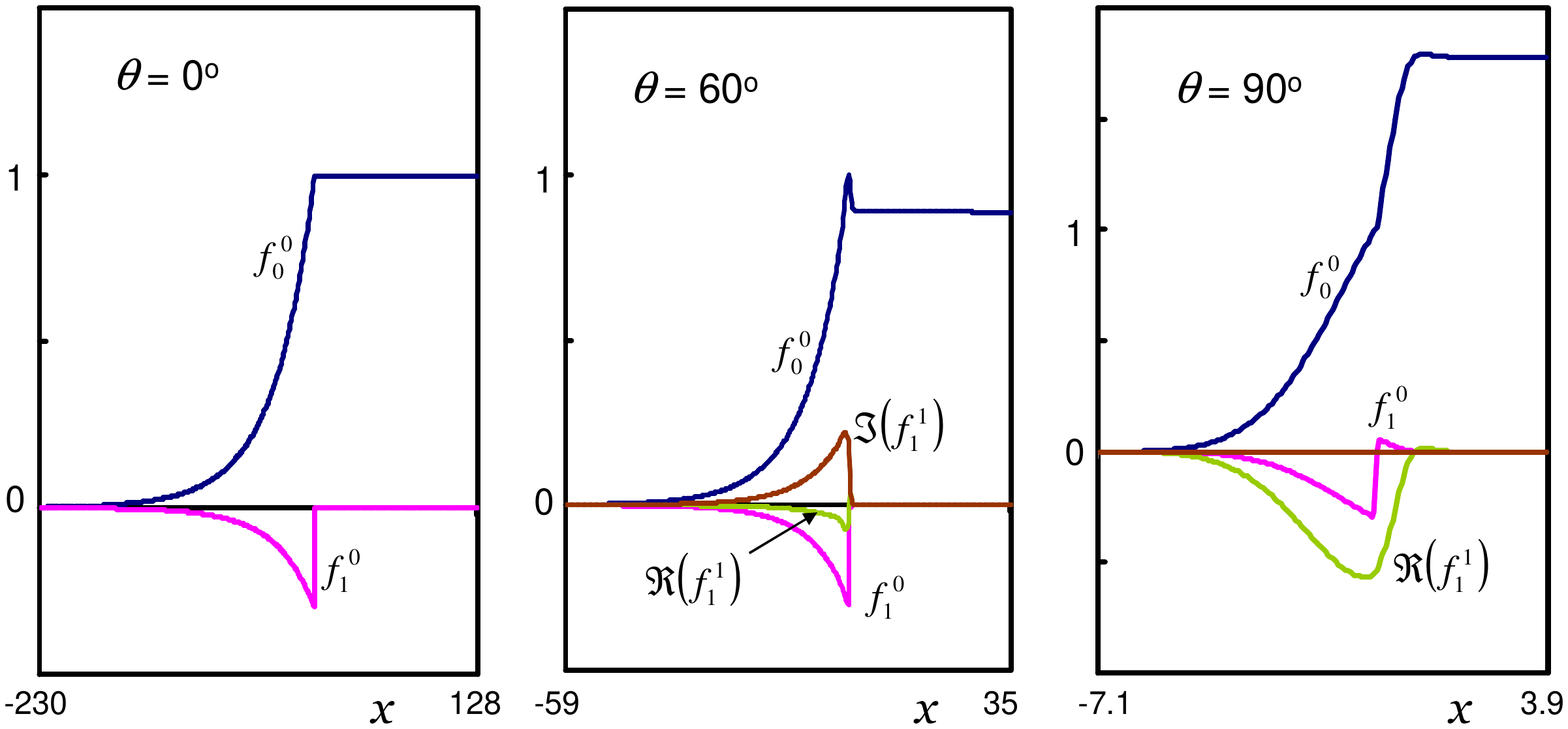}
\vskip 2 cm 
\caption{Spatial profiles of $f_0^0$, $f_1^0$, 
$\Re\{f_1^1\}$, and
$\Im\{f_1^1\}$ for a parallel shock, an oblique shock at $\theta=60^\circ$ and a peperpendicular shock.  
$\nu/\omega _g=0.1$ and $u_s=c/10$.
The spatial distance $x$ from the shock at $x=0$ is
measured in CR Larmor radii.  The vertical axis is scaled such that $f_0^0=1$ at the shock. }
\label{fig:Fig2}
\end{center}
\end{figure*}

Figure 2 displays profiles of $f_0^0$, $f_1^0$ and $f_1^1$ for
a parallel shock ($\theta =0^\circ$),
an oblique shock ($\theta =60^\circ$),
and a perpendicular shock ($\theta =90^\circ$).
Figure 3 shows the detailed profiles in the immediate environment of the shock
for $\theta =60^\circ,\ 72^\circ,\  \&\  90^\circ $.
In figures 2 and 3, $\nu/\omega _g=0.1$ and $u_s=c/10$.
The oblique shock ($\theta =60^\circ$) exhibits a peak in the CR density at the shock as 
previously identified in mildly relativistic shocks.  
Geissler et al (1999)
convincingly argued for a build-up of CR immediately upstream of the shock
because CR approaching the shock from upstream with certain pitch angles
are unable to cross the shock into the downstream region.
Similar structures are evident in Monte Carlo calculations by Ellison et al (1996).
Figure 3 is a plot of CR density profiles at higher resolution.
At $\theta = 60 ^\circ$ an excess of CR is seen within one CR Larmor radius both
upstream and downstream of the shock.
According to equations (7) and (8), as discussed above, the excess CR density at the shock
results in enhanced CR acceleration.  The excess increases the rate at which CR cross the shock, and hence increases the probability of acceleration to high energy before being advected away downstream of the shock.  This explains the flattening of the CR spectrum at quasi-parallel shocks as seen in figure 1.

\begin{figure*}
\begin{center}
\includegraphics[angle=0,width=9cm]{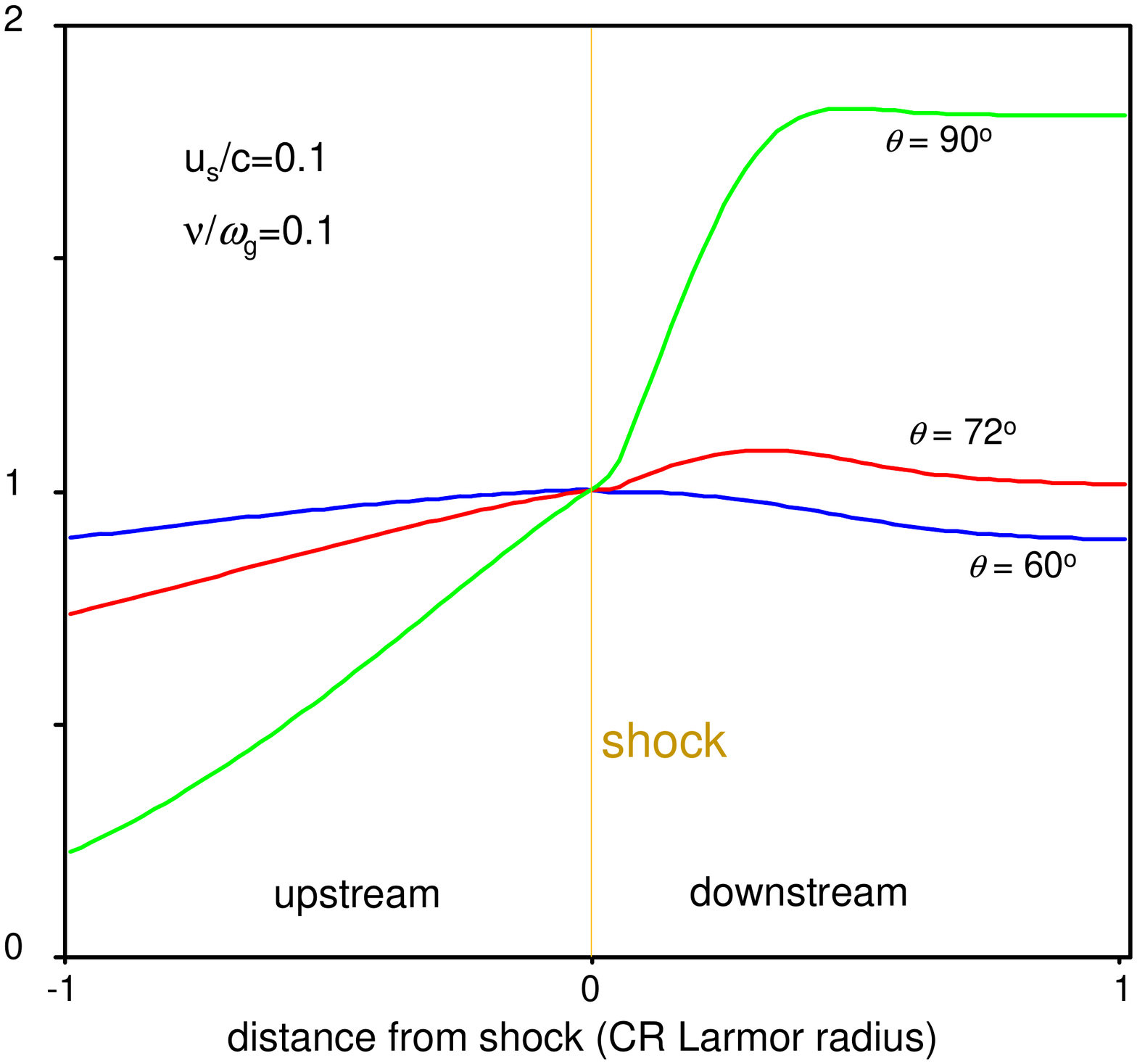}
\vskip 2 cm 
\caption{Spatial profiles of the CR density $f_0^0$ within one CR Larmor radius either
side of the shock.  $\nu/\omega _g=0.1$ and $u_s=c/10$.}
\label{fig:Fig3}
\end{center}
\end{figure*}

The CR density profile at a perpendicular shock is different as seen in Figure 2(c) and in figure 3.
Diffusion theory can still be applied to CR at a perpendicular shock
by deconstructing the CR trajectory into diffusive motion of the
gyrocentre and gyration about the gyrocentre.
The gyratory part of the trajectory acts as a spatial convolution on the
motion of the gyrocentre.  This spreads the CR trajectory 
over a distance of one Larmor radius and smooths the CR density at the shock.
A discontinuous jump in CR density is not possible at the shock on a scale less than a Larmor radius.
Correspondingly, figures 2(c) and 3 exhibit a CR density gradient both downstream and upstream
of a perpendicular shock.
Consequently, the CR density precisely at the shock is smaller
than the CR density far downstream. 
In this case, the rate at which CR cross the shock is reduced
and the spectrum at quasi-perpendicular shocks is steeper than $p^{-4}$ in line with equations (7) and (8).
According to equations (9) and (10) of Jokipii (1987) the diffusive approximation is only valid at a quasi-perpendicular shock
when $\nu/\omega _g > u_s/c$.  This is consistent with the results shown in our figure 1
where we see that deviations from a $p^{-4}$ spectrum occur where this condition is not met.

The profile for $\theta = 72^\circ$ is intermediate between the
quasi-perpendicular and the quasi-parallel cases.
The steepening and flattening effects are both present but cancel out
to leave a CR density approximately equal to that far downstream
and a CR spectrum proportional to $p^{-4}$.


\section{Observations}

{\it (i) Galactic cosmic rays}

CR arriving at the earth with energies up to a few PeV conform to a remarkably straight
power law spectrum $p^{-\gamma}$ with $\gamma$ equal to 4.6-4.7.
The more rapid escape from the galaxy of high energy CR steepens the spectrum, 
but this cannot be solely responsible for steepening the spectrum
from an index at source of $\gamma=4.0$.  
The underlying CR source spectrum probably has an
index between about 4.2 and 4.3 (Gaisser, Protheroe \& Stanev 1998, Hillas 2005).
Figure 1 indicates that the oblique-shock effects discussed here can produce a steepened 
spectrum ($\gamma >4$) at high shock velocities,
but only if the shock has a tendency towards a quasi-perpendicular rather than
quasi-parallel configuration.  As discussed in section 4, the spectral index averaged
over random magnetic field orientations is approximately the same as for the low
velocity limit: $\gamma=4$.

The shock might be
preferentially quasi-perpendicular in the event of strong magnetic field amplification
in the precursor due to CR streaming (Bell 2004, 2005).  
Field amplification is driven by ${\bf j}_{CR}\times {\bf B}$ where ${\bf j}_{CR}$ is the electric
current carried by streaming CR and ${\bf B}$ is the local magnetic field.  The plasma
is set in motion perpendicular to ${\bf j_{CR}}$ and therefore predominantly
perpendicular to the shock normal since ${\bf j_{CR}}$ is directed away from the shock. 
These upstream fluid motions stretch the magnetic field
perpendicular to the shock normal producing a quasi-perpendicular shock orientation
as evident in the sample field lines shown in figure 4 of Bell (2005).
Hence strong field amplification can be expected to lead preferentially to a quasi-perpendicular
shock geometry and the steepening of the spectral index. 
This is not contradicted by polarimetric observations which suggest a predominantly radial magnetic field in SNR.
Zirakashvili \& Ptuskin (2008) point out that this field orientation probably develops downstream of the shock
due to differential fluid motions after the plasma has passed through the shock.
A similar effect is seen in simulations by Giacalone \& Jokipii (2007).
Alternatively, the shock might
be preferentially quasi-perpendicular if the supernova expands into a pre-existing
circumstellar wind supporting a magnetic field in the form of a Parker spiral.
This would be unsurprising since SN are often preceded by strong mass loss.

Another possible source of spectral steepening is non-linear feedback onto the shock
hydrodynamics (Drury \& V\"{o}lk 1981, Drury 1983).
Non-linear effects produce a concave spectrum which is steepened at low energy but flatter than $p^{-4}$
at high energy (Bell  1987, Falle \& Giddings 1987).
Allen et al (2008) find evidence for concavity in the synchrotron spectrum of SN1006.
The galactic CR spectrum is not noticeably concave, but a compensating convexity due to
momentum-dependent escape from the galaxy could produce the measured spectrum.
The CR spectrum could be the result of a complicated mixture of non-linear, oblique-shock and escape effects.

\vskip 0.3 cm
\noindent{\it (ii)  Young supernova remnants}

Further evidence for deviation from a $p^{-4}$ CR source spectrum can be found in synchrotron spectra.
SNR radio spectral indices can deviate considerably from
$\alpha=0.5$ ($\gamma=4$). 
Since synchrotron loss times at radio wavelengths are longer than the dynamical time, 
the measured spectral index gives a proper indication of the cosmic ray source spectrum. 
Observations show that young pre-Sedov SNR
consistently exhibit radio spectral
indices significantly steeper than $\alpha =0.5$.
Either magnetic field amplification or expansion into a Parker spiral may
produce a geometry favouring quasi-perpendicular shocks and spectral steepening.
But in either case, a high shock velocity is needed.  If this is the explanation
of the steeper than expected galactic CR spectrum, then it implies that galactic
CR are accelerated by young SNR rather than by SNR in the Sedov phase.

In Figure 4 we plot a sample of SNR (see Table~\ref{table:spectralindex}) 
and their spectral indices as a function of shock velocity. It is related to the analysis of Glushak (1985) 
who plotted the spectral indices of young shell SNR against SNR age to show that the spectra flattened with time. 
Here we plot the spectral index
against the {\em mean} expansion velocity $v_{\rm sh}$ calculated as the mean radius divided
by SNR age, and extend and update the sample of SNR. We take the mean rather than the momentary expansion velocity 
since the overall spectrum is an addition of cosmic rays that were accelerated throughout the entire age of the remnant. 
Depending on the number of cosmic rays that enroll in the acceleration process early versus late, 
additional differences in the spectral index can result. 
The average time since acceleration would be different for core-collapse SNR and Type Ia SNe.  
In Type Ia SNe most CR are accelerated at late times as the shock expands into a uniform medium, 
whereas CR acceleration is more evenly spread in the case of core-collapse SNR 
which expand into a $r^{-2}$ density profile of a circumstellar wind (Schure et al 2010).

The Galactic remnants in figure 4 were taken from the updated online catalogue by Green (2009). 
They were selected if they were classified as having a clear shell morphology (i.e. no contamination in radio 
from a central source), had a measured spectral index and a readily available distance estimate. 
Only sources were selected that are estimated to be younger than 10,000~yr, 
and not clearly interacting or overlapping with other radio sources.

Because of a lack of data for galactic SNR with ages less than 200 years, we
turn to the extragalactic remnants of recent SN for data on very young SNR.  
We added data for SN1987A (Manchester et al 2002, Manchester et al 2005) and SN1993J (Weiler et al 2007), 
and a sample of Type Ib/c supernovae taken from Chevalier \& Fransson (2006). %
Two data points are plotted for SN1987A corresponding to observations in 2002 
and 2005 since its expansion velocity slowed significantly between these years. 
 
For our purpose, very young SNR from Type Ib/c SNe are especially interesting because of their high expansion velocities.
They do not necessarily reflect a statistical sample. 
Their expansion velocities are based on models of the radius of the remnant at the time of the 
peak in synchrotron self absorption, as described in (Chevalier 1998).  
Each of the Type Ib/c's has
a steep spectral index except for SN1998bw which was the only one to be associated with a GRB. 
Chevalier \& Fransson (2006) argue that the flattening with time may be a result of the nonlinear 
feedback of the cosmic rays in the plasma, which weakens with time.  
Nonlinear models by Ellison \& Reynolds (1991) can partly account for the steepening, 
up to $\alpha\approx 0.8$, if the cosmic ray pressure constitutes a dominant fraction. 
The total steepening may be a cumulative effect of non-linear and oblique-shock steepening.

Except in the cases of the historical supernova remnants, the age of the galactic remnants is a highly uncertain parameter. 
Often it is based on the distance measurement, giving the size of the remnant, 
and from the temperature in X-rays a shock velocity is derived. Subsequently a Sedov model 
is used to arrive at the age estimate. 
However, in addition to uncertainties in the size, 
if any energy is lost in e.g.~cosmic rays the shock velocity may be severely underestimated, 
as was shown might be the case in RCW 86 (Helder et al 2009). 
The data points in Figure 4 should therefore be considered with large error bars in $v_{\rm sh}$ 
and are likely to underestimate the shock velocity on average. 
The ones for which this is known are indicated with an arrow, 
however this does not mean that other data points are necessarily more certain. 
Occasionally the shock velocity can possibly be overestimated when the age measurement 
is based on the age of the associated pulsar (i.e.~G292.2-0.5). 
The age of the historical remnants is obviously more certain. 
However, their distance estimate may still introduce a large uncertainty in the size and shock velocity.
Nevertheless, the data show a clear trend. 
As a guide, we plot on figure 4 a trend-line:
$$\alpha=0.7+0.3 \log_{10}(v_{\rm sh}/10^4 {\rm km \ s}^{-1})$$
where $v_{\rm sh}$ is the radius divided by age for SNR. 
Although the data are uncertain at high velocity, the rise in the spectral index levels off at about $\alpha =1$. 
A levelling off also occurs at $\alpha \sim 0.5$ for older SNR with lower expansion velocities.
A better fit to the overall data may be an S-shaped curve with a stronger increase of spectral index at
velocities around $10^4$km s$^{-1}$.  
An appropriate trend-line
for the historical SNR, for which the velocity estimates are more reliable, is 
$$\alpha=0.7+0.8 \log_{10}(v_{\rm sh}/10^4 {\rm km \ s}^{-1})$$
as plotted in figure 4.

The two data points for SN1987A show that the variation of spectral index with shock velocity 
is exhibited within the lifetime of a single SNR.
The radio spectrum of Cassiopeia A has similarly been observed to flatten with time as 
already argued in the seventies (Baars et al 1977).
A flattening of the Cas A radio spectrum from $\alpha=0.79$ to $\alpha=0.77$ was found over a period of 15 years (1965--1980). 
Further flattening was confirmed by O'Sullivan \& Green (1990), although they found it proceeding at a slower rate.

\vskip 0.3 cm
\noindent{\it (iii)  Spectral index variations within SNR}

Further signatures of spectral index variation with shock obliquity may be
visible in SNR expanding at high velocity into a uniform magnetic field.  
If magnetic field amplification is weak, the field orientation at the shock 
is that of the pre-existing interstellar field.  
The electron spectrum should be flatter
where the shock is quasi-parallel, increasing the number of CR electrons accelerated to high energy, and
producing stronger X-ray synchrotron emission at the poles.

Detailed, resolved, spectra are available for the remnant of SN1006. 
This image shows a distinct bipolar symmetry, which is thought to reflect the directionality of the ambient magnetic field. 
Most likely the bright caps are the parts where the shock velocity is parallelly aligned 
with the background magnetic field (Rothenflug et al 2004, Bocchino et al
2011). In an analysis of the combined radio- with the X-ray spectrum Allen et al (2008) 
found evidence for hardening of the spectrum in this area, both for models which included curvature in the spectrum 
and those without. The fractional hardening is modest but significant. 
The magnetic field in SN 1006 is generally believed to be not much amplified and therefore plausibly reflects the 
original field geometry. The spectral steepening towards the fainter parts, 
i.e.~higher obliquity angle between the shock normal and magnetic field lines, is consistent with our findings. 

Detailed data are also available for Tycho's SNR, as was presented in Katz-Stone (2000). 
This remnant shows a much more patchy profile in radio. Spectra were extracted from various parts over the remnant. 
No clear trend with position was found. 
However, it was argued that the brightest parts have a flatter spectral index than the steeper parts, 
which is a similar trend to that in SN1006. 

A spatially resolved study of the radio spectrum of Cassiopeia A (Wright et al 1999) shows variation of the 
spectral index with position in the remnant. Generally, a steeper power law is found in the outermost regions, 
whereas in the inner bright ring the spectral index is flatter (although still $\alpha >0.75$). 
The spectra they found do not show signs of spectral curvature and seem to be coherent over larger scales 
than the diffusion scale. This indicates a common mechanism, such as e.g.~a preferentially perpendicular field 
at the forward shock might be able to account for. 
The magnetic field in Cas A is strongly amplified and thus a preferentially perpendicular field may be expected. 

\vskip 0.3 cm
\noindent{\it (iv)  Older supernova remnants}

The radio spectral indices of old SNR in the self-similar Sedov phase are usually around 0.5, 
but many have flatter spectra which may be due to 
additional second order Fermi acceleration in MHD turbulence (Ostrowski 1999),
or possibly due to compression ratios greater than four at radiative shocks.
Because the shock velocities are much lower than in young SNR, the oblique-shock effects considered in this paper are
unlikely to contribute to the flattening of the spectral index.

\begin{figure*}
\begin{center}

\includegraphics[width=\textwidth]{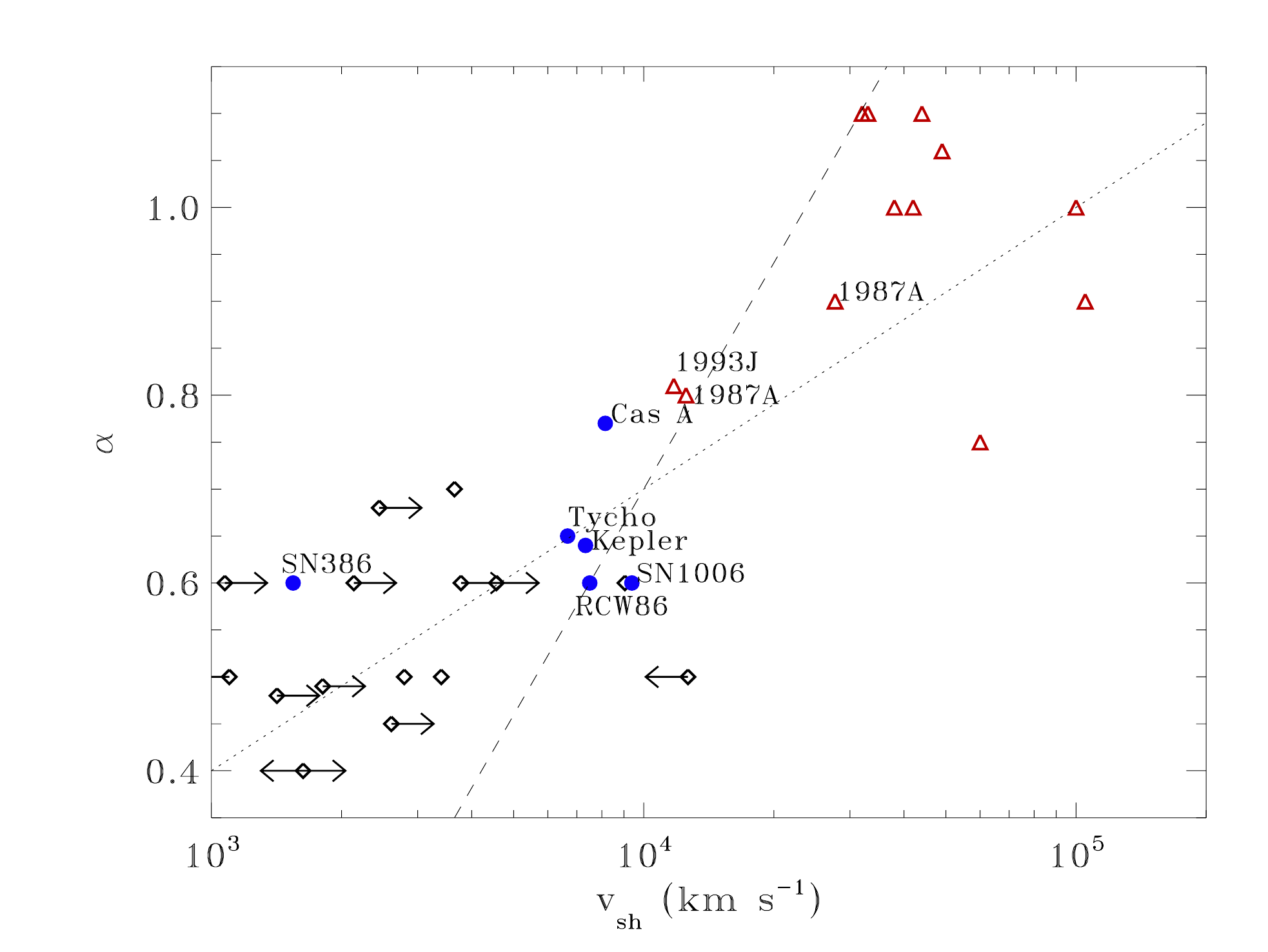}
\caption{Synchrotron spectral index versus shock velocity for galactic (black diamonds), historic (blue filled circles), 
and extragalactic SNR and SNe (red triangles), as given in Table~1. 
The arrows indicate known issues with distance and / or age estimates 
and the likely limit derived from that in shock velocity. 
Note that there remain large errors in $v_{\rm sh}$ also for the other data points. 
The dashed line shows the trend in spectral index that is biased towards the historical galactic remnants 
$\alpha=0.7+0.8 \log_{10}(v_{\rm sh}/10^4 {\rm km \ s}^{-1})$. 
The dotted line shows a more average trend: $\alpha=0.7+0.3 \log_{10}(v_{\rm sh}/10^4 {\rm km \ s}^{-1})$. 
}
\label{fig:fitsnrdata}
\end{center}
\end{figure*}
\section{Conclusions}

We have shown that a wide range of spectral indices are possible for shock
velocities of around $c/30$ or higher.  The CR spectrum
is steepened for quasi-perpendicular shocks and flattened for
shocks closer to parallel geometry.  
This may present a natural explanation of why few observed energetic particle spectra 
conform to the non-relativistic test-particle $\gamma=4$ spectrum, although
non-linear effects and second order Fermi acceleration probably also play a role.
For a shock propagating into a randomly orientated upstream magnetic field
the steepening and flattening of the spectrum averages out to produce $\gamma =4$.
Observed radio spectra show that the spectrum is steeper at high shock velocity which
suggests that high velocity shocks tend to be quasi-perpendicular rather than parallel.
A predominantly quasi-perpendicular configuration may arise if magnetic field
amplification is strong as expected at high shock velocity.
Alternatively, a quasi-perpendicular geometry may result from 
expansion into a circumstellar wind supporting a Parker spiral.

The galactic CR energy spectrum and the radio spectra of young SNR require a CR spectrum at source
which is steeper than $p^{-4}$.
The galactic CR spectral index may be explained if CR are accelerated by
shocks such as those found in the historical SNR.  
These SNR have a relatively large shock area, still expand at a relatively high velocity 
into a possibly dense circumstellar medium, and
produce CR with a suitable spectral index if the SNR shock is predominantly quasi-perpendicular.
The spectral index is sensitive to the CR scattering rate $\nu$
which is presently not well determined by observation or theory.
Observations (figure 4) suggest a fairly clear relation between 
SNR shock velocity and spectral index.

Our model depends on the presumed simplification that the magnetic field structure
can be separated into a small scale disordered component that scatters CR
through small angles,  
and a large scale component which is uniform over a CR Larmor radius.
However, the essential reason for spectral steepening by quasi-perpendicular shocks is that
the magnetic field sweeps CR through the shock making it difficult for 
CR to return to the shock for further acceleration.
This effect can be expected to be independent of the detailed field structure.
Another realisation of a related effect would be spectral steepening due to
field line wandering (Kirk, Duffy \& Gallant 1996).

\section{Acknowledgements}

The research leading to these results has received funding
from the European Research Council under the European
Community's Seventh Framework Programme (FP7/2007-
2013) / ERC grant agreement no. 247039 and from grant number ST/H001948/1
made by the UK Science Technology and Facilities Council.
We especially thank John Kirk of MPIK Heidelberg for very helpful
discussions.

\newpage

\section{References}
Achterberg A., Gallant Y.A.  Kirk J.G. \&  Guthmann A.W., 2001, MNRAS 328, 393
\newline
Allen G.~E., Houck J.~C. \& Sturner S.~J., 2008,
ApJ, 683, 773
\newline
Axford W.I., Leer E., Skadron G., 1977, Proc 15th Int. Cosmic Ray Conf., 11, 132
\newline
Baars J.~W.~M., Genzel R., Pauliny-Toth I.~I.~K. \&  Witzel A., 1977,
A \& A, 61, 99
\newline
Baring M.G., Ellison D.C. \& Jones F.C., 1993, ApJ, 409, 327
\newline
Bell A.R., 1978, MNRAS, 182, 147
\newline
Bell A.R. 1987 MNRAS, 225, 615
\newline
Bell A.R., 2004, MNRAS, 353, 550
\newline
Bell A.R., 2005, MNRAS, 358, 181
\newline
Bell A.R., Robinson A.P.L., Sherlock M. Kingham R.J. \& Rozmus W.,
2006, Plasma Phys Cont Fusion, 48, R37
\newline
Blandford R.D. \&  Ostriker J.P., 1978, ApJ, 221, L29
\newline
Bocchino F., Orlando S., Miceli M. \& Petruk O., 2011, A\&A 539, 129
\newline
Chen Y., Su Y., Slane P.~O. \& Wang Q.~D., 2004,
ApJ, 616, 885
\newline
Chevalier R.~A., 1998,
ApJ, 499, 810
\newline
Chevalier R.~A. \&  Fransson C., 2006,
ApJ, 651, 381
\newline
Chiotellis A., Schure K.~M., \& Vink J., 2011 arXiv 1103.5487
\newline
Crawford F., Gaensler B.~M., Kaspi V.~M., Manchester R.~N., Camilo F., Lyne A.~G. \& Pivovaroff M.~J. 2001,
ApJ, 554, 152
\newline
Drury L.O'C., 1983, Rep Prog Phys, 46, 973
\newline
Drury L.O'C. \& V\"{o}lk H.J., 1981, ApJ 248, 344
\newline
Ellison D.C., Baring  M.G \& Jones F.C., 1996, ApJ 473, 1029
\newline
Ellison D.~C. \& Reynolds S.~P., 1991,
ApJ, 382, 242
\newline
Falle S.A.E.G. \& Giddings J.R. 1987 MNRAS 225, 399
\newline
Gaensler B.~M., Manchester R.~N. \& Green A.~J., 1998,
MNRAS, 296, 813
\newline
Gaisser TK, Protheroe RJ \& Stanev T, 1998, ApJ 492, 219
\newline
Giacalone, J. \& Jokipii, J.R., 2007, ApJ 663, L41
\newline
Gieseler U.D.J., Kirk J.G., Gallant Y.A. \& Achterberg A., 1999, A\&A 345 298
\newline
Glushak A.P., 1985, Pis'ma Astron Zh  11, 825
\newline
G{\'o}mez Y., \& Rodr{\'{\i}}guez L.~F. 2009 Rev Mex Astron Astrophys 45, 91
\newline
Gotthelf E.~V., Petre R. \& Hwang U., 1997,
ApJL, 487, L175
\newline
Green D.~A., 2009,  http://www.mrao.cam.ac.uk/surveys/snrs/
\newline
Greiner J., Egger R. \& Aschenbach B., 1994,
A \& A, 286, L35
\newline
Harrus I.~M. \& Slane P.~O., 1999,
ApJ, 516, 811
\newline
Haslam C.~G.~T., Salter C.~J. \& Pauls T., 1980,
A \& A, 92, 57
\newline
Helder E.~A., Vink J., Bassa C.~G., Bamba A., 
Bleeker J.~A.~M., Funk S., Ghavamian P., van der Heyden K.~J., Verbunt F. \& Yamazaki R.,
2009, Science 325, 719
\newline
Hillas A.M., 2005, J Phys G 31, R95
\newline
Huba J.D., 1994, NRL Plasma Formulary, publ. Naval Research Laboratory, Washington, USA
\newline
Hwang U., Petre R. \& Hughes J.~P., 2000,
ApJ, 532, 970
\newline
Johnston T.W., 1960, Phys Rev 120, 1103
\newline 
Jokipii J.R., 1982, ApJ 215, 716
\newline
Jokipii J.R., 1987, ApJ 313, 842
\newline
Katz-Stone D.~M., Kassim N.~E. Lazio T.~J.~W. \& O'Donnell R., 2000,
ApJ, 529, 453
\newline
Kinugasa K. Torii K. Tsunemi H., Yamauchi S., Koyama K. \& Dotani T., 1998,
Publ.~Astron.~Soc.~Japan, 50, 249
\newline
Kirk J.G., Duffy P. \& Gallant Y.A., 1996, A\&A 314, 1010
\newline
Kirk J.G. \& Heavens A.F., 1989, MNRAS 239, 995
\newline
Kobayakawa K., Honda Y.S. \& Samura T., 2002, Phys Rev D 66 083004
\newline
Krymskii G.F., 1977, Sov Phys Dokl, 23, 327
\newline
Manchester R.~N., Gaensler B.~M., Staveley-Smith L., Kesteven M.~J. \& Tzioumis A.~K., 2005,
ApJL, 628, L131
\newline
Manchester R.~N., Gaensler B.~M., Wheaton V.~C., Staveley-Smith L., 
Tzioumis A.~K., Bizunok N.~S., Kesteven M.~J.  \& Reynolds J.~E.,
2002,
Pub.~Astr.~Soc.~Austr., 19, 207
\newline
Ostrowski M., 1991, MNRAS 249 551
\newline
Ostrowski M., 1999, A\&A, 345, 256
\newline
O'Sullivan C. \& Green D.~A., 1999,
MNRAS, 303, 575
\newline
Rakowski C.~E., Hughes J.~P.\&  Slane, P., 2001,
ApJ, 548, 258
\newline
Rothenflug R., Balle, J., Dubner G., Giacani E., Decourchelle A. \& Ferrando, P., 2004, A\&A, 425, 121
\newline
Ruffalo D., 1999, ApJ 515 787
\newline
Sankrit R., Blair W.~P., Delaney T., Rudnick L., Harrus I.~M.\&  Ennis J.~A., 2005,
Advances in Space Research, 35, 1027
\newline
Schure K.~M., Achterberg A., Keppens R. \& Vink J., 2010, MNRAS 406, 2633
\newline
Slane P., Chen Y., Lazendic J.~S. \& Hughes J.~P., 2002,
ApJ, 580, 904
\newline
Tian W.~W., Leahy D.~A., 2008,
MNRAS, 391, L54
\newline
Weiler K.~W., Williams C.~L., Panagia N., Stockdale C.~J., Kelley M.~T., Sramek R.~A., Van Dyk S.~D. \& Marcaide J.~M.,
2007, ApJ, 671, 1959
\newline
Wright M., Dickel J., Koralesky B. \& Rudnick L., 1999, 
ApJ, 518, 284
\newline
Zhou X., Chen Y., Su Y. \& Yang J., 2009, ApJ, 691, 516
\newline 
Zirakashvili V.N. \& Ptuskin V.S., 2008, ApJ 678, 939
\newpage

\begin{table}

\centering                 
\begin{tabular}{c c  c   c  c  }       
\hline\hline
SNR & diameter & age & velocity (R/t) &$\alpha$ 
\\
& pc & years & $10^3$~km~s$^{-1}$ &  \\
\hline
G1.9+0.3 & 3.7 & 220$^b$ & 9 & 0.6 \\
G4.5+6.8 (Kepler) & 6$^c$ & 400 & 7.3 & 0.64 \\
G11.2-0.3 (SN 386) & 5.1 & 1620 & 1.5 & 0.6 \\
G21.8-0.6 (Kes 69) & 32 & 4600$^d$ & 3.4 &0.5 \\
G27.4+0.0 (Kes 73) & 10 & 1000$^e$ & 4.9 & 0.68 \\
G29.7-0.3 (Kes 75) & 6.6 & 884 & 3.7 & 0.7 \\
G31.9+0.0 (3C391) & 14.8 & 4000$^f$ & 1.8 & 0.49 \\
G39.2-0.3 (3C396) & 15.5 & 7100$^g$ & 1.1 & 0.6 \\
G43.3-0.2 (W49B) & 11.6 & 4000$^h$ & 1.4 & 0.48 \\
G93.3+6.9 (DA 530) & 16 & 3000$^i$ & 2.6 & 0.45 \\
G111.7-2.1 (Cas A) & 5 & 330 & 7.4 & 0.77 \\
G120.1+1.4 (Tycho) & 6 & 440 & 6.7 & 0.65 \\
G272.2-3.2 & 7.9$^j$ & 1100 & 3.5 & 0.6 \\
G292.2-0.5 & 44 & 1700$^k$ & 13 & 0.5 \\
G296.8-0.3 & 56 & 6000$^l$ & 4.6 & 0.6 \\
G315.4-2.3 (RCW86) & 28 & 1825 & 7.5 & 0.6 \\
G327.6+14.6 (SN1006) & 19.2 & 1000 & 9.4 & 0.6 \\
G332.4-0.4 (RCW103) & 9 & 4000$^m$ & 4.4 & 0.5 \\
G337.2-0.7 & 10 & 3000$^n$ & 1.6 & 0.4 \\
G349.7+0.2 & 16 & 2800$^p$ & 2.8 & 0.5 \\
G352.7-0.1 & 17 & 2200$^q$ & 3.1 & 0.6 \\
SN1987A$^r$ & 0.34 & 6 & 35 & 0.9 \\
SN1987A$^s$ & 0.41 & 16 & 10 & 0.8 \\
SN1993J$^t$ & 0.24 & 10 & 15 & 0.81 \\
SN1983N$^v$ & & & 42 & 1.0 \\
SN1984L$^v$ & & & 100 & 1.0 \\
SN1990B$^v$ & & & 33 & 1.0 \\
SN1994I$^v$ & & & 38 & 1.0 \\
SN2001ig$^v$ & & & 49 & 1.06 \\
SN2002ap$^v$ & & & 105 & 0.9 \\
SN2003L$^v$ & & & 32 & 1.1 \\
SN2003bg$^v$ & & & 44 & 1.1 \\
SN1998bw$^v$ & & & 60 & 0.75 \\
\\
\hline                                   
\end{tabular}
\caption{\textrm{Overview of the supernova remnant models.
Data are taken from the updated Galactic SNR catalogue by Green (2009) unless other source specified. 
Other sources:
$^b$Gomez \& Rodriguez (2009), 
$^c$Sankrit et al (2005), Chiotellis et al (2011), 
$^d$Zhou et al (2009), 
$^e$Tian \& Leahy (2008), 
$^f$Chen et al (2004), 
$^g$Harrus \& Slane (1999), 
$^h$Hwang et al (2000), 
$^i$Haslam et al (1980), 
$^j$Greiner et al (1994), 
$^k$Crawford et al (2001), 
$^l$Gaensler et al (1998), 
$^m$Gotthelf et al (1997),
$^n$Rakowski et al (2001), 
$^p$Slane et al}, 
$^q$Kinugasa et al (1998),
$^r$Manchester et al (2002), 
$^s$Manchester et al (2005), 
$^t$Weiler et al (2007), 
$^v$Chevalier \& Fransson (2006) 
} 
\label{table:spectralindex}   
\end{table}


\label{lastpage}
\end{document}